\documentclass[twocolumn,showpacs,preprintnumbers,amsmath,amssymb]{revtex4}

\usepackage{graphicx}
\usepackage{bm}
\usepackage{color}
\usepackage{amsmath}
\usepackage{amsfonts}
\usepackage{amssymb}
\usepackage{ulem}

\begin{document}

\preprint{APS/123-QED}

\title{Coupled spin-1/2 antiferromagnetic chain Cs$_2$LiRuCl$_6$ with partially disordered crystal lattice}


\author{Kazumitsu~Watanabe$^1$}
\author{Nobuyuki~Kurita$^1$}
\author{Hidekazu~Tanaka$^1$}
\email{tanaka@lee.phys.titech.ac.jp}
\author{Haruki~Sugiyama$^2$}
\author{Hidehiro~Uekusa$^3$}
\author{Ismael F. Diaz-Ortega$^4$}
\author{Hiroyuki~Nojiri$^4$}

\affiliation{
$^1$Department of Physics, Tokyo Institute of Technology, Meguro-ku, Tokyo 152-8551, Japan\\
$^2$Research and Education Center for Natural Sciences, Keio University, Kohoku-ku, Yokohama 223-8521, Japan\\
$^3$Department of Chemistry, Tokyo Institute of Technology, Meguro-ku, Tokyo 152-8551, Japan\\
$^4$Institute for Materials Research, Tohoku University, Aoba-ku, Sendai 980-8577, Japan
}

\date{\today}

\begin{abstract}

We determined the crystal structure of Cs$_2$LiRuCl$_6$, which was synthesized first in this work, and investigated its magnetic properties. Cs$_2$LiRuCl$_6$ has a hexagonal structure composed of linear chains of face-sharing RuCl$_6$ and LiCl$_6$ octahedra. In two-thirds of the structural chains, Ru$^{3+}$ and Li$^+$ sites are almost ordered, while in the other chains their sites are disordered. This situation is analogous to the ground state of the antiferromagnetic Ising model on a triangular lattice. Using electron paramagnetic resonance, we evaluated the $g$ factors of Ru$^{3+}$ with effective spin-1/2 as $g_c\,{=}\,2.72$ and $g_{ab}\,{=}\,1.50$ for magnetic fields $H$ parallel and perpendicular to the $c$ axis, respectively. Magnetization curves for $H\,{\parallel}\,c$ and $H\,{\perp}\,c$ are highly anisotropic. However, these magnetization curves approximately coincide when normalized by the $g$ factors. It was found from the magnetization and specific heat results that Cs$_2$LiRuCl$_6$ can be described as a coupled one-dimensional $S\,{=}\,1/2$ Heisenberg-like antiferromagnet with $J/k_{\rm B}\,{\simeq}\,3.7$ K. Three-dimensional ordering occurs at $T_{\rm N}\,{=}\,0.48$ K. A magnetic phase diagram for $H\,{\parallel}\,c$ is also presented.

\end{abstract}


\maketitle

\section{Introduction\label{intro}}

The honeycomb-lattice Kitaev model is one of the hot topics in condensed matter physics~\cite{Kitaev,Motome}. The Kitaev model is expressed by the Ising model $-J^{\gamma}S_i^{\gamma}S_j^{\gamma}$ for three different spin components $S_i^{\gamma}$ (${\gamma}\,{=}\,x, y$ and $z$) on three different links. It was demonstrated that this spin model can be described by itinerant and localized Majorana fermions with very different energy scales and that the ground state is exactly a quantum spin liquid state~\cite{Kitaev}. 
Experimental studies of the Kitaev model were stimulated by the theoretical prediction that this model can be realized when MX$_6$ octahedra centered by magnetic M ions such as Ru$^{3+}$ and Ir$^{4+}$ with effective spin-1/2 owing to the strong spin-orbit coupling are linked by sharing their edges to form a honeycomb lattice~\cite{Jackeli}.  A$_2$IrO$_3$ (A\,{=}\,Li, Na and Cu)~\cite{Singh,Singh2, Mehlawa,Liu,Choi,Ye,Williams,Choi2,Takahashi} and $\alpha$-RuCl$_3$~\cite{Plumb,Kubota,Sears,Johnson2,Cao,Banerjee3,Hirobe,Wolter,Ponomaryov,Banerjee,Baek,Do, Banerjee2,Yamauchi,Kasahara,Nagai,Sears2,Ran,Widmann}, which approximately satisfy such a structural condition, have been actively investigated via various experimental techniques. Their ground states were found to be not the quantum spin liquid state but ordered states, which are considered to be caused by the presence of the Heisenberg term~\cite{Liu, Choi,Ye,Chaloupka,Jeffrey,Sears,Johnson2,Banerjee,Cao,Ran,Ponomaryov,Sears2}.

For $\alpha$-RuCl$_3$, three-dimensional ordering occurs at $T_{\rm N}\,{=}\,7.6$ K~\cite{Kubota,Do,Widmann,Sears,Johnson2,Cao,Sears2,Ran,Ponomaryov}, which is considered to be caused by the additional Heisenberg term~\cite{Janssen}. However, static and dynamic properties characteristic of the system of Majorana fermions, such as the two-stage temperature structure of entropy~\cite{Kubota,Do,Widmann} and the intense excitation continuum near the ${\Gamma}$ point~\cite{Banerjee3,Do,Banerjee}, have been observed. The zigzag magnetic ordering~\cite{Johnson2,Ran, Cao} is strongly suppressed by the external magnetic field applied parallel to the honeycomb layer~\cite{Kubota,Wolter,Banerjee2,Sears2,Johnson2}. Consequently, the ordered ground state changes into the disordered state~\cite{Baek,Banerjee2,Nagai}, in which the half-integer quantization of thermal Hall conductance is observed as predicted for the Kitaev model~\cite{Kasahara}. 

Theory predicts that in a triangular lattice, the competition between the Kitaev and Heisenberg terms produce an exotic helical order called a $Z_2$ vortex crystal~\cite{Kimchi,Becker,Catuneanu,Shinjo,Rousochatzakis,Kos,Kishimoto,Perkins}. The helical order is different from the conventional helical order that arises from the competition between two or more exchange interactions of the Heisenberg type or between the exchange interaction and the Dzyaloshinskii-Moriya interaction. Magnetic insulators composed of Ru$^{3+}$ and Ir$^{4+}$ are expected to exhibit various ground states, and thus, new compounds are desired. With this in mind, we synthesized Cs$_2$LiRuCl$_6$. In this paper, we report the crystal structure of this compound and its low-temperature magnetic properties.

\begin{figure}[t]
\centering
\includegraphics[width=8cm,clip]{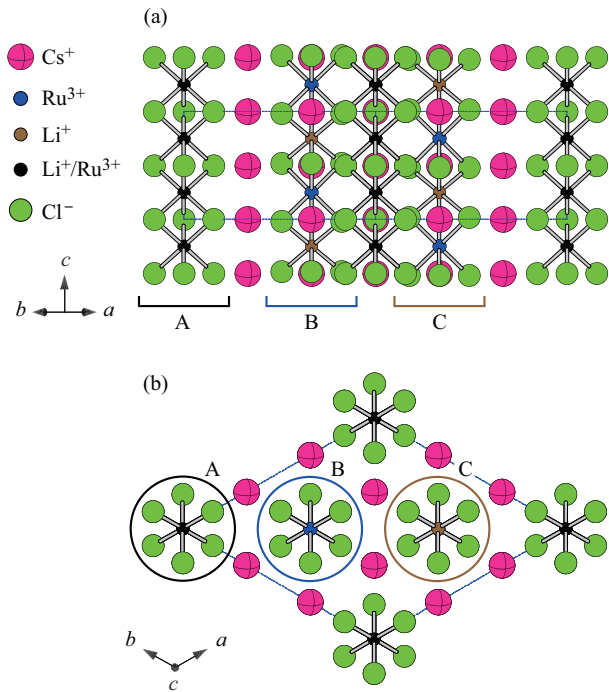}
\caption{(Color online) Crystal structure of Cs$_2$LiRuCl$_6$ viewed along the (a) $[1, 1, 0]$, and (b) $c$ directions. Dotted blue lines denote the chemical unit cell. The crystal structure consists of three kinds of chain, A, B and C, composed of face-sharing RuCl$_6$ and LiCl$_6$ octahedra, which are enclosed by black, blue and ocher circles, respectively. In chain A, RuCl$_6$ and LiCl$_6$ octahedra are randomly distributed with a probability of 1/2, while in chains B and C, they are arranged almost alternately. The atomic arrangement in chain C is obtained by shifting that in chain B by $c/2$.
}
\label{fig:Structure}
\end{figure}

\section{Experimental details\label{experimental}}
To prepare Cs$_2$LiRuCl$_6$ single crystals, we first prepared Cs$_3$Ru$_2$Cl$_9$ crystals by the vertical Bridgman method from a melt comprising a stoichiometric mixture of CsCl and RuCl$_3$ sealed in an evacuated quartz tube. A mixture of Cs$_3$Ru$_2$Cl$_9$, CsCl and LiCl in a molar ratio of $1\,{:}\,1\,{:}\,2$ was vacuum-sealed in a quartz tube. The temperature at the center of the furnace was set at 900$^{\circ}$C, and the lowering rate was 3 mm$\cdot$h$^{-1}$. Single crystals of about 1 cm$^3$ were obtained. These crystals were found to be Cs$_2$LiRuCl$_6$ from X-ray diffraction, as shown in Section \ref{structure}. The crystals are easily cleaved parallel to the $c$ axis.

The specific heat was measured at temperatures down to 0.35\,K in magnetic fields of up to 9\,T using a physical property measurement system (PPMS, Quantum Design) by a relaxation method. 
The magnetization was measured down to $T\,{=}\,0.5$\,K and up to $H\,{=}\,7$\,T 
using a SQUID magnetometer (MPMS-XL, Quantum Design) equipped with a $^3$He device (iHelium3, IQUANTUM).
High-magnetic-field magnetization measurement was performed at the Institute for Materials Research (IMR), Tohoku University. The temperature of the sample was lowered to 0.4 K using liquid ${}^3$He. A magnetic field of up to 20 T was applied with a multilayer pulse magnet.
High-magnetic-field electron spin resonance (ESR) measurement was also conducted in the frequency range of $110\,{-}\,360$ GHz and the temperature range of $4.2\,{-}\,50$ K at IMR, Tohoku University. Gunn oscillators were used as light sources.

\section{Crystal Structure\label{structure}}
Because the crystal structure of Cs$_2$LiRuCl$_6$ has not been reported to date, we performed its structural analysis at 293 and 93 K using a RIGAKU R-AXIS RAPID three-circle X-ray diffractometer equipped with an imaging plate area detector. Monochromatic Mo-K$\alpha$ radiation with a wavelength of ${\lambda}\,{=}\,0.71075$\,\rm{\AA} was used as the X-ray source. Data integration and global cell refinements were performed using data in the range of $3.28^{\circ}\,{<}\,{\theta}\,{<}\,27.52^{\circ}$, and absorption correction was performed using the ABSCOR program~\cite{Higashi}. The total number of reflections observed was 7842, among which 625 reflections were found to be independent and 495 reflections were determined to satisfy the criterion $I\,{>}\,2{\sigma}(I)$. Structural parameters were refined by the full-matrix least-squares method using SHELXL$-$2018/8 software~\cite{Sheldrick}.  The final $R$ indices obtained for $I\,{>}\,2{\sigma}(I)$ were $R\,{=}\,0.0464$ and $wR\,{=}\,0.1033$. The crystal data are listed in Table \ref{table:1}.  The chemical formula was confirmed to be Cs$_2$LiRuCl$_6$. The crystal structure of Cs$_2$LiRuCl$_6$ is hexagonal $P6_322$ with cell dimensions of $a\,{=}\,12.4194(16)$\,$\rm{\AA}$, $c\,{=}\,6.0621(7)$\,$\rm{\AA}$ and $Z\,{=}\,3$. Its atomic coordinates, equivalent isotropic displacement parameters and site occupancies are shown in Table \ref{table:2}. 

\begin{table}
\caption{Crystal data for Cs$_2$LiRuCl$_6$ at 293 K.}
\label{table:1}
\begin{center}
\begin{tabular}{cccc}
\hline
& Chemical formula & Cs$_2$LiRuCl$_6$ &  \\
& Space group & $P6_322$ &  \\
& $a$ ($\rm{\AA}$) & 12.4194(16) &  \\
& $c$ ($\rm{\AA}$) & 6.0621(7) &  \\
& $V$ ($\rm{\AA}^3$) & 809.8(2) &  \\
& $Z$ & 3 &\\
& $R;\ wR$ &  0.0464;\ 0.1033 & \\
\hline
\end{tabular}
\end{center}
\end{table}

\begin{table}
\caption{Fractional atomic coordinates, equivalent isotropic displacement parameters and site occupancy for Cs$_2$LiRuCl$_6$ at 293 K.}
\label{table:2}
\begin{tabular}{rrrrrr}
\hline
Atom  &  $x$\hspace{7mm}    & $y$\hspace{7mm}   & $z$\hspace{5mm}  & $U_{\rm eq}$\hspace{1mm} & occ.  \\ \hline
Cs  &  0.000000  &  0.6666(3)  &  0.500000  &  0.0336(4) & 1 \\
Li(1)  &  0.000000  &  0.000000  &  0.250000  &  0.0189(12) & 0.5 \\
Li(2)  &  0.666667  &  	0.333333  &  0.750000  &  0.035(5) & 0.922(6) \\
Li(3)  &  0.666667  &  	0.333333  &  0.250000  &  0.0167(4) & 0.078(6)\\
Ru(1)  &  0.000000 &  0.000000 &  0.250000  &  0.0189(12) & 0.5 \\
Ru(2)  &  0.666667 &  0.333333 &  0.750000 &  0.035(5) & 0.078(6) \\
Ru(3)  &  0.666667 &  0.333333 &  0.250000 &  0.0167(4) & 0.922(6) \\
Cl(1)  &  0.1557(6)  &  0.000000 &  	0.500000 &  0.0333(16) & 1 \\
Cl(2)  &  0.6658(7)  &  0.4885(5) &  0.4816(2) &  0.0268(8) & 1 \\
\hline
\end{tabular}
\end{table}

We also conducted the structural analysis at $T\,{=}\,93$ K and confirmed that the crystal structure is the same as that determined at 293 K. The lattice constants at $T\,{=}\,93$ K are $a\,{=}\,12.3582(8)$\,$\rm{\AA}$ and $c\,{=}\,6.0321(4)$\,$\rm{\AA}$.

The crystal structure viewed along the $[1, 1, 0]$ and $c$ directions is illustrated in Fig.\,\ref{fig:Structure}. The crystal structure consists of three kinds of chain, A, B and C, twhich are composed of face-sharing RuCl$_6$ and LiCl$_6$ octahedra. In chain A, RuCl$_6$ and LiCl$_6$ octahedra are randomly distributed with a probability of 1/2, while in chains B and C, they are arranged almost alternately in order. The atomic arrangement in chain C is obtained by shifting that in chain B by $c/2$, and thus, chains B and C are equivalent. The crystal structure of Cs$_2$LiRuCl$_6$ is different from the Cs$_2$LiGaF$_6$ structure~\cite{Babel}, although Rb$_2$LiRuCl$_6$ was reported to be isostructural with Cs$_2$LiGaF$_6$~\cite{Meyer}. 
All the RuCl$_6$ octahedra in Cs$_2$LiRuCl$_6$ are trigonally elongated along the $c$ axis in contrast to the case of $\alpha$-RuCl$_3$, where all the RuCl$_6$ octahedra are trigonally compressed along the $c$ axis. This is consistent with the experimental result that in Cs$_2$LiRuCl$_6$, the $g$ factor $g_c$ for a magnetic field parallel to the $c$ axis is greater than $g_{ab}$ for a magnetic field parallel to the $ab$ plane, as shown in the next Section. 

\begin{figure}[b]
\centering
\includegraphics[width=6.5cm,clip]{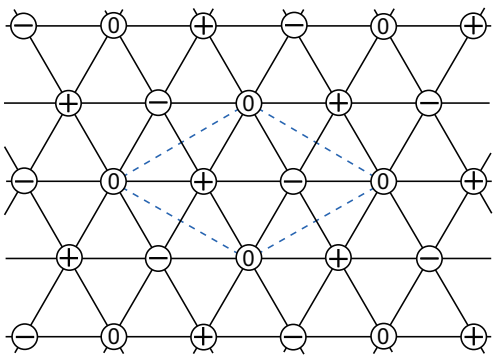}
\caption{(Color online) Ground state of an antiferromagnetic Ising model on a triangular lattice with the nearest-neighbor interaction with finite entropy~\cite{Wannier}. Notations ``$+$" and ``$-$" denote ordered sites with $\langle{\sigma}^z\rangle\,{=}\,{\pm}1$, respectively, and ``0" denotes the disordered site with $\langle{\sigma}^z\rangle\,{=}\,0$. Dashed lines show the magnetic unit cell.}
\label{fig:Ising}
\end{figure}

Ru$^{3+}$ and Li$^+$ ions occupy the lattice points of a triangular lattice in the $ab$ plane. The arrangement of both ions is analogous to the ground state of an antiferromagnetic Ising model on a triangular lattice (AFIMTL) with the nearest-neighbor (NN) interaction, as shown in Fig.~\ref{fig:Ising}~\cite{Wannier}. Using the Ising spin ${\sigma}^z\,{=}\,{\pm}1$, lattice points occupied by Ru$^{3+}$ and Li$^+$ are made to correspond to ${\sigma}^z\,{=}\,+1$ and $-1$, respectively. If the total energy is lowered when different ions occupy the neighboring sites, the arrangement of Ru$^{3+}$ and Li$^+$ ions can be mapped onto the AFIMTL with the NN interaction. Figure~\ref{fig:Ising} shows the ground state of the AFIMTLT with finite entropy~\cite{Wannier}. In the ordered ``$+$" and ``$-$" sites, spins are fixed to ${\sigma}^z\,{=}\,+1$ and $-1$, respectively, while in the ``0" site, the spin is disordered to be $\langle{\sigma}^z\rangle\,{=}\,0$. This is because the mean fields acting on the ``0" site cancel out. 
This spin state is called a partially disordered state~\cite{Mekata}. If the disordered site is regarded as the site occupied by Ru$^{3+}$ and Li$^+$ ions with an equal probability, the ordering pattern of Fig.~\ref{fig:Ising} is equivalent to the arrangement of Ru$^{3+}$ and Li$^+$ ions shown in Fig.~\ref{fig:Structure}(b). Thus, it is likely that the site disorder of Ru$^{3+}$ and Li$^+$ ions in structural chain A arises from crystallographic frustration.

\section{Magnetic properties\label{results}}

\begin{figure}[t]
\centering
\includegraphics[width=8cm,clip]{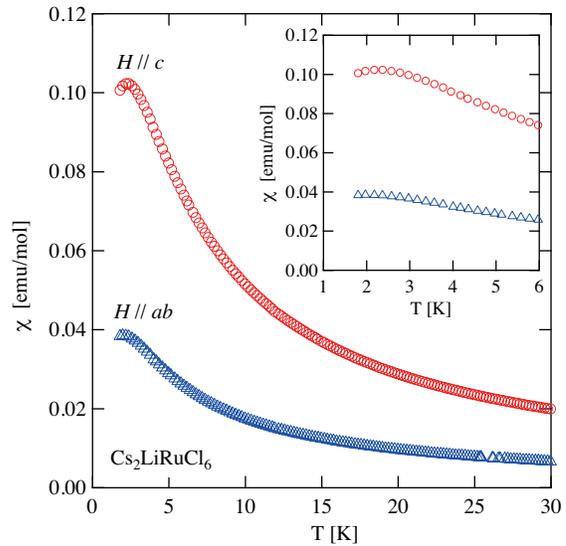}
\caption{(Color online) Temperature dependences of magnetic susceptibilities $\chi$ measured for $H\,{\parallel}\,c$ and $H\,{\parallel}\,ab$ at $H\,{=}\,0.1$ T. The inset shows the enlargement of magnetic susceptibilities below 6 K.}
\label{fig:sus}
\end{figure}

Figure~\ref{fig:sus} shows the temperature dependences of the magnetic susceptibilities ${\chi(T)}$ of Cs$_2$LiRuCl$_6$ measured for magnetic fields $H$ parallel to the $c$ axis ($H\,{\parallel}\,c$) and the $ab$ plane ($H\,{\parallel}\,ab$). The absolute values of the susceptibilities are highly anisotropic. This mainly arises from the anisotropy of the $g$ factor, as shown below. The susceptibility has a rounded maximum at $T_{\rm max}({\chi})\,{\simeq}\,2.2$ K. This is suggestive of the low dimensionality of the exchange network in Cs$_2$LiRuCl$_6$.  

\begin{figure}[ht]
\centering
\includegraphics[width=8.0cm,clip]{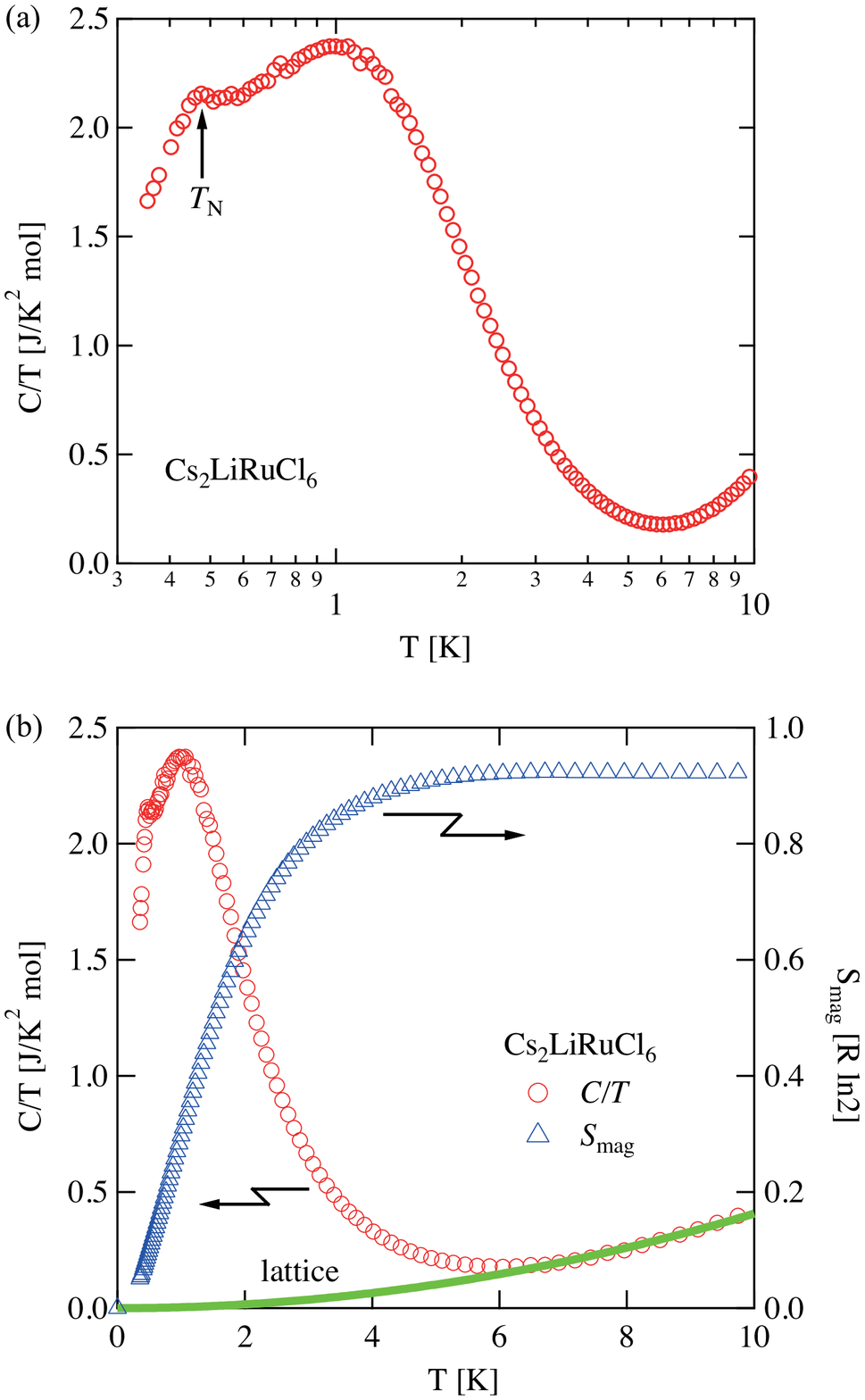}
\caption{(Color online) (a) Total specific heat divided by the temperature $C/T$ as a function of logarithmic temperature measured at zero magnetic field. The vertical arrow indicates the magnetic phase transition temperature $T_{\rm N}\,{=}\,0.48$ K. (b) Temperature dependences of $C/T$ and magnetic entropy measured at zero magnetic field. The green solid line is the lattice contribution.}
\label{fig:heat0}
\end{figure}

Figure~\ref{fig:heat0}(a) shows the total specific heat divided by the temperature $C/T$ of Cs$_2$LiRuCl$_6$ as a function of logarithmic temperature measured at zero magnetic field. With decreasing temperature, $C/T$ exhibits a rounded maximum at $T_{\rm max}(C/T)\,{\simeq}\,1$ K. With further decreasing temperature, $C/T$ displays a cusp anomaly at $T_{\rm N}\,{=}\,0.48$ K indicative of a magnetic phase transition.

Figure~\ref{fig:heat0}(b) shows the temperature dependences of the total specific heat divided by the temperature $C/T$ and the magnetic entropy $S_{\rm mag}$ of Cs$_2$LiRuCl$_6$ at zero magnetic field. The green solid line is the lattice contribution $C_{\rm lat}/T$ estimated by assuming $C_{\rm lat}/T\,{\propto}\,T^2$. The magnetic entropy saturates to $S_{\rm mag}\,{=}\,0.92\,R\,{\ln}\,2$ above 7 K. This indicates that Ru$^{3+}$ in Cs$_2$LiRuCl$_6$ is in the low-spin state with the effective spin-1/2.

\begin{figure}[ht]
\centering
\includegraphics[width=8.5cm,clip]{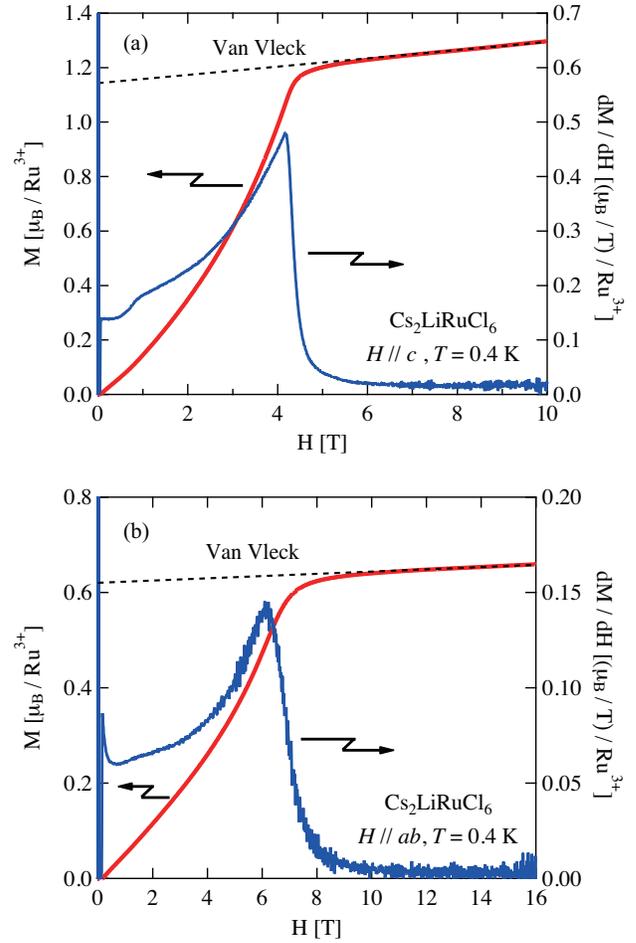}
\caption{(Color online) Magnetic field dependence of the raw magnetization $M$ (left) and its field derivative $dM/dH$ (right) measured at $T\,{=}\,0.4$ K for (a) $H\,{\parallel}\,c$ and (b) $H\,{\parallel}\,ab$. Dashed lines denote the Van Vleck paramagnetism.}
\label{fig:mag1}
\end{figure}

Figure~\ref{fig:mag1} shows magnetization curves $M(H)$ of Cs$_2$LiRuCl$_6$ and their field derivatives $dM/dH$ measured for $H\,{\parallel}\,c$ and $H\,{\parallel}\,ab$ at $T\,{=}\,0.4$ K. Dashed lines denote the Van Vleck paramagnetism. The Van Vleck paramagnetic susceptibilities for $H\,{\parallel}\,c$ and $H\,{\parallel}\,ab$ were evaluated as ${\chi}_{\rm VV}^c\,{=}\,8.49\,{\times}\,10^{-3}$\,emu/mol and ${\chi}_{\rm VV}^{ab}\,{=}\,1.31\,{\times}\,10^{-3}$\,emu/mol, respectively. 

A kink anomaly is observed in $dM/dH$ for $H\,{\parallel}\,c$ at $H_{\rm c}\,{\simeq}\,0.7$ T as shown in Fig.~\ref{fig:mag1}(a). This magnetization anomaly is indicative of a spin-flop-like transition in the ordered state because the temperature is lower than $T_{\rm N}\,{=}\,0.48$ K.

\begin{figure}[ht]
\centering
\includegraphics[width=8.5cm,clip]{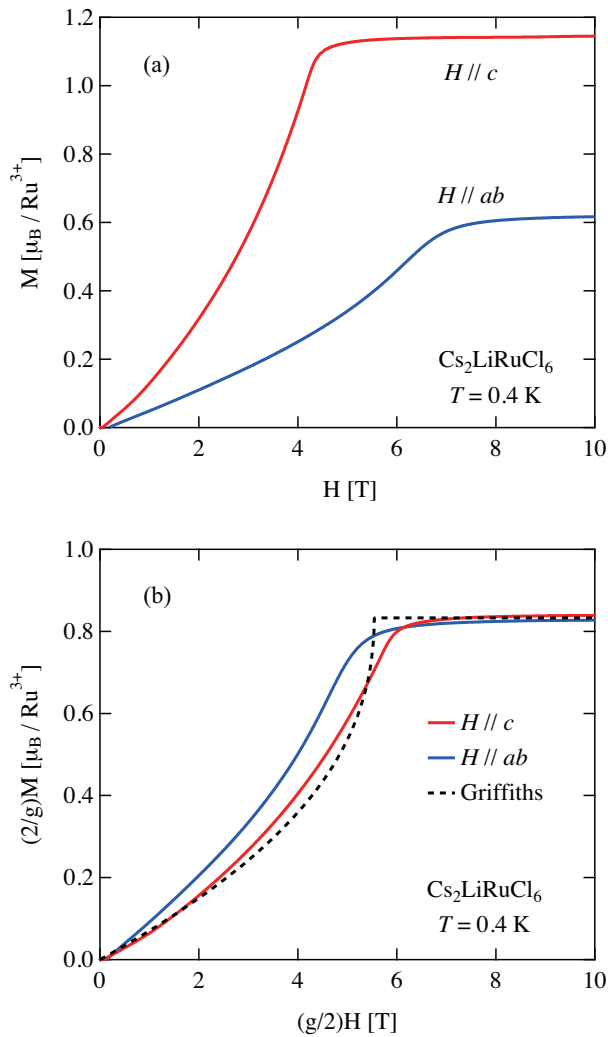}
\caption{(Color online) (a) Magnetization curves at $T\,{=}\,0.4$ K for $H\,{\parallel}\,c$ and $H\,{\parallel}\,ab$ corrected for the Van Vleck paramagnetism. (b) Magnetization curves at $T\,{=}\,0.4$ K for $H\,{\parallel}\,c$ and $H\,{\parallel}\,ab$ normalized by the $g$ factors $g_c\,{=}\,2.72$ and $g_{ab}\,{=}\,1.50$. The dashed line is the magnetization curve for the spin-1/2 antiferromagnetic Heisenberg chain calculated with $J/k_{\rm B}\,{=}\,3.7$ K~\cite{Griffiths}.}
\label{fig:mag2}
\end{figure}

Figure~\ref{fig:mag2}(a) shows the magnetization curves at $T\,{=}\,0.4$ K for $H\,{\parallel}\,c$ and $H\,{\parallel}\,ab$ corrected for the Van Vleck paramagnetism. The magnetization curves are highly anisotropic, which is mainly attributed to the anisotropy of the $g$ factor, as shown below. The saturation field and magnetization are $H^c_{\rm s}\,{=}\,4.35$ T and $ M^c_{\rm s}\,{=}\,1.14$ ${\mu}_{\rm B}$/Ru$^{3+}$ for $H\,{\parallel}\,c$, and $H^{ab}_{\rm s}\,{=}\,6.90$ T and $ M^{ab}_{\rm s}\,{=}\,0.62$ ${\mu}_{\rm B}$/Ru$^{3+}$ for $H\,{\parallel}\,ab$, respectively.

Figure~\ref{fig:mag2}(b) shows the magnetization curves for $H\,{\parallel}\,c$ and $H\,{\parallel}\,ab$ normalized by the $g$ factors $g_c\,{=}\,2.72$ and $g_{ab}\,{=}\,1.50$, which were obtained by the present ESR measurement. The normalized values of the saturation magnetization $(2/g)M_{\rm s}$ for $H\,{\parallel}\,c$ and $H\,{\parallel}\,ab$ are in good agreement, although the normalized saturation fields $(g/2)H_{\rm s}$ are different. The difference in normalized saturation field can be attributed to the anisotropy of the exchange interaction. 

The magnetization anomaly at saturation for $H\,{\parallel}\,c$ is reasonably sharp, thus, the saturation field is well defined, although the saturation anomaly for $H\,{\parallel}\,ab$ is rather smeared, which is ascribed to the magnetic anisotropy such as the Dzyaloshinskii-Moriya interaction and the Kitaev term. This is in contrast to the case of $\alpha$-RuCl$_3$, where the saturation of magnetization is considerably smeared owing to the large Kitaev term, which does not commute with the total spin~\cite{Kubota,Johnson2}. The well-defined saturation anomaly of magnetization in Cs$_2$LiRuCl$_6$ indicates that the Kitaev term is relatively small as compared with the Heisenberg term.

The saturation magnetization normalized by the $g$-factor is $(2/g)M_{\rm s}\,{=}\,0.84$, which is smaller than unity. As shown in Section \ref{structure}, RuCl$_6$ and LiCl$_6$ octahedra are randomly arranged on average in chain A. When two RuCl$_6$ octahedra are adjacent, sharing their face similarly to Ru$_2$Cl$_9$ double octahedra, the exchange interaction should be strongly antiferromagnetic and its magnitude is estimated to be $J/k_{\rm B}\,{\simeq}\,700-900$ K from the exchange constant in Cs$_3$Ru$_2$Cl$_9$. This compound is composed of face-sharing Ru$_2$Cl$_9$ double octahedra, which form a magnetic dimer~\cite{Lockwood}. The exchange constant in the dimer is as large as $J/k_{\rm B}\,{\simeq}\,700-900$ K. Therefore, the contribution of two adjacent RuCl$_6$ octahedra to magnetization is considered to be negligible. If half of the RuCl$_6$ octahedra are adjacent, sharing their face, and the other half form a ${-}$\,RuCl$_3$\,${-}$\,LiCl$_3$\,${-}$ chain, then the normalized saturation magnetization is evaluated to be $(2/g)M_{\rm s}\,{=}\,0.83$, which is close to the experimental value of $(2/g)M_{\rm s}\,{=}\,0.84$.

\begin{figure}[t]
\centering
\includegraphics[width=8cm,clip]{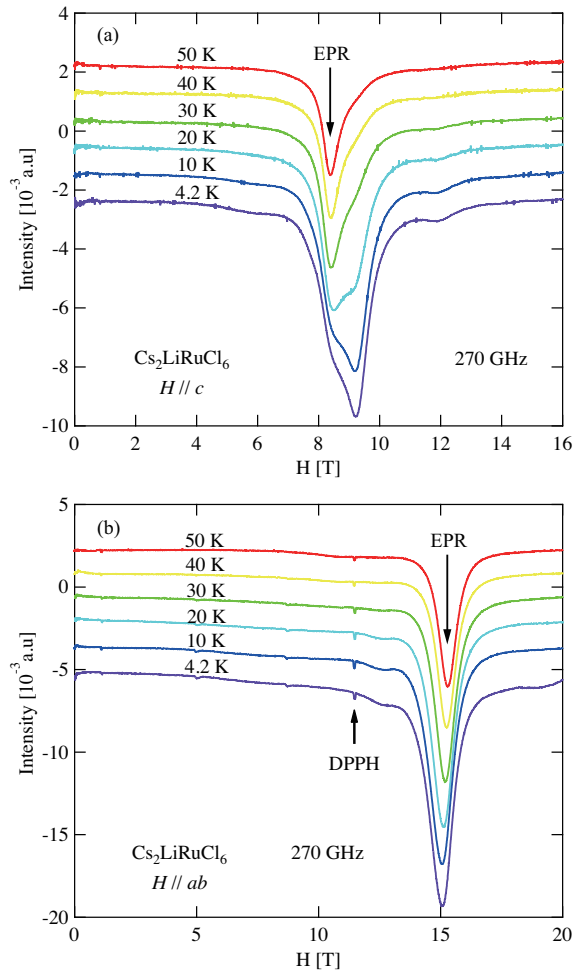}
\caption{(Color online) ESR spectra measured at ${\nu}\,{=}\,270$ GHz and at various temperatures for (a) $H\,{\parallel}\,c$ and (b) $H\,{\parallel}\,ab$. Small sharp lines labeled DPPH denote $g$\,=\,2. Resonance signals labeled EPR denote the electron paramagnetic resonance measured at $T\,{=}\,50$ K.}
\label{fig:spectra}
\end{figure}


Figure~\ref{fig:spectra} shows ESR spectra measured at ${\nu}\,{=}\,270$ GHz and at various temperatures for $H\,{\parallel}\,c$ and $H\,{\parallel}\,ab$. For $H\,{\parallel}\,ab$, a single electron paramagnetic resonance (EPR) peak is observed near $H\,{=}\,15$ T at all temperatures, while for $H\,{\parallel}\,c$, a new resonance peak appears on the high-field side below 40 K. Because the resonance field is higher than the saturation field, the new resonance mode is an ESR mode in the forced ferromagnetic state. At present, the origin of the new resonance mode is unclear because the spin structure in the ordered state has not been solved. 

\begin{figure}[ht]
\centering
\includegraphics[width=8cm,clip]{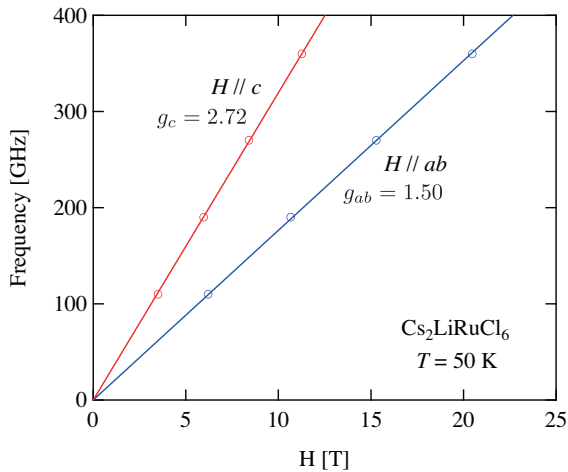}
\caption{(Color online) Frequency-field diagram of the EPR lines at $T\,{=}\,50$ K for $H\,{\parallel}\,c$ and $H\,{\parallel}\,ab$. From the slopes of these lines, the $g$ factors for $H\,{\parallel}\,c$ and $H\,{\parallel}\,ab$ are determined to be $g_c\,{=}\,2.72$ and $g_{ab}\,{=}\,1.50$, respectively.}
\label{fig:diagram}
\end{figure}

We obtained the $g$ factor from the EPR line measured at $T\,{=}\,50$ K. Figure~\ref{fig:diagram} shows the frequency-field diagram of the EPR lines for $H\,{\parallel}\,c$ and $H\,{\parallel}\,ab$ measured at $T\,{=}\,50$ K. We can see that the resonance fields are exactly proportional to the frequency. From the slopes of the EPR lines, the $g$ factors for $H\,{\parallel}\,c$ and $H\,{\parallel}\,ab$ are obtained as $g_c\,{=}\,2.72$ and $g_{ab}\,{=}\,1.50$, respectively. The condition of $g_c\,{>}\,g_{ab}$ in Cs$_2$LiRuCl$_6$ is consistent with the trigonally elongated RuCl$_6$ octahedron, as shown below.

\begin{figure}[t]
\centering
\includegraphics[width=8cm,clip]{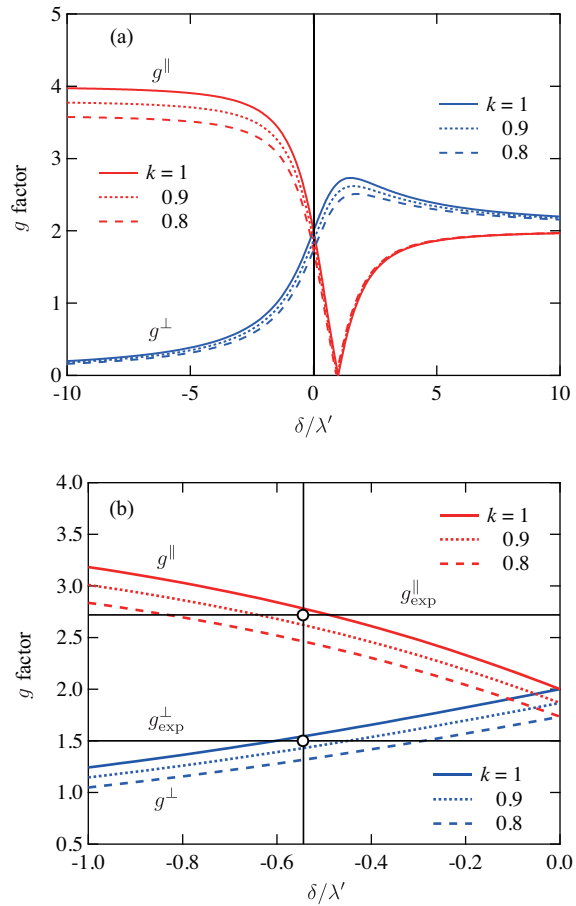}
\caption{(Color online) (a) $g$ factors as a function of ${\delta}/{\lambda}^{\prime}$ calculated for three reduction factors, $k\,{=}\,1.0, 0.9$ and 0.8, where $g^{\parallel}$ and $g^{\perp}$ are $g$ factors for magnetic fields parallel and perpendicular to the trigonal axis of a RuCl$_6$ octahedron, respectively. (b) Enlargement of the $g$ factors between ${\delta}/{\lambda}^{\prime}\,{=}\,-1$ and 0. The two horizontal lines are experimental $g$ factors determined by the present ESR measurement. Open circles represent a set of $g$ factors suitable for Cs$_2$LiRuCl$_6$. }
\label{fig:g_factor}
\end{figure}

We next discuss the $g$ factor of Ru$^{3+}$ in trigonal crystalline field following Ref.~\cite{Kubota}. In the low-spin state of Ru$^{3+}$, all five electrons in the $4d$ orbitals occupy the $d{\epsilon}$ orbital. The orbital state is triply degenerate. The orbital degeneracy can be lifted by the spin-orbit coupling and the trigonal crystalline field, which are written using the orbital angular momentum ${\bm l}$ with $l\,{=}\,1$ as 
\begin{eqnarray}
{\cal H}^{\prime}={\lambda}^{\prime}({\bm l}\cdot{\bm S})+{\delta}\left\{(l^z)^2-2/3\right\},
\label{perturbation}
\end{eqnarray}
where ${\lambda}^{\prime}=k{\lambda}$ and the second term represents the energy of the trigonal crystalline field. ${\lambda}$ is the coupling constant of the spin-orbit coupling and $k$ ($0\,{<}\,k\,{\leq}\,1$) is the reduction factor, which expresses the reduction of the matrix elements of the angular momentum owing to the mixing of the $p$ orbital of the surrounding Cl$^-$ with the $4d$ orbitals of Ru$^{3+}$. When the RuCl$_6$ octahedron is trigonally compressed, ${\delta}>0$, and when it is elongated, ${\delta}<0$. 

The orbital triplet splits into three Kramers doublets. When the temperature $T$ is much lower than ${\lambda}^{\prime}\,{\simeq}\,1000$~cm$^{-1}$~\cite{Geschwind}, i.e., $T\,{<}~100$\,K, the magnetic property is determined by the lowest Kramers doublet. Its eigenvalue is expressed as
\begin{eqnarray}
\frac{E_0}{{\lambda}^{\prime}}=-\frac{{\delta}}{6{\lambda}^{\prime}}-\frac{1}{4}-\frac{1}{2}\sqrt{\left(\frac{{\delta}}{{\lambda}^{\prime}}\right)^2-\frac{{\delta}}{{\lambda}^{\prime}}+\frac{9}{4}}.
\label{eigen}
\end{eqnarray}
The eigenstates of the lowest Kramers doublet are expressed as
\begin{eqnarray}
{\psi}_{\pm}=c_1|\pm 1, \mp 1/2{\rangle}+c_2|0, \pm 1/2{\rangle},
\label{state}
\end{eqnarray}
where $|m_l, m_S{\rangle}$ denotes the state with $l^z\,{=}\,m_l$ and $S^z\,{=}\,m_S$. Coefficients $c_1$ and $c_2$ are given by
\begin{eqnarray}
c_1=\frac{1}{\sqrt{2}}\sqrt{1\,{-}\,\frac{A}{\sqrt{A^2{+}1}}},\ \ 
c_2=-\frac{1}{\sqrt{2}}\sqrt{1\,{+}\,\frac{A}{\sqrt{A^2{+}1}}}
\label{coefficient}
\end{eqnarray}
with
\begin{eqnarray}
A=\frac{2({\delta}/{\lambda}^{\prime})-1}{2\sqrt{2}}.
\label{A}
\end{eqnarray}
The lowest Kramers doublet splits into two Zeeman levels when subjected to a magnetic field. The splitting of the Zeeman levels is proportional to the $g$ factor, which is expressed as 
\begin{eqnarray}
g^{\parallel}=2|\{(k+1)c_1^2-c_2^2\}|
\label{g_para}
\end{eqnarray}
for a magnetic field parallel to the trigonal axis, and
\begin{eqnarray}
g^{\perp}=2(c_2^2-\sqrt{2}kc_1c_2)
\label{g_perp}
\end{eqnarray}
for a magnetic field perpendicular to the trigonal axis. Figure~\ref{fig:g_factor} shows these $g$ factors as a function of ${\delta}/{\lambda}^{\prime}$. 

When a RuCl$_6$ octahedron is trigonally elongated, $g^{\parallel}\,{>}\,g^{\perp}$, and when it is compressed, $g^{\parallel}\,{<}\,g^{\perp}$. In Cs$_2$LiRuCl$_6$, the $g$ factors obtained for magnetic fields parallel and perpendicular to the crystallographic hexagonal axis were $g_c\,{=}\,2.72$ and $g_{ab}\,{=}\,1.50$, respectively. These $g$ factors are consistent with the elongated RuCl$_6$ octahedron determined by the present X-ray diffraction experiment. 
Figure~\ref{fig:g_factor}(b) shows the behavior of the $g$ factors in the range of $-1\,{\leq}\,{\delta}/{\lambda}^{\prime}\,{\leq}\,0$. There is one set of parameters, $({\delta}/{\lambda}^{\prime}, k)\,{=}\,(-0.54, 0.96)$, that satisfies the $g$ factors observed in Cs$_2$LiRuCl$_6$. The fact that the large anisotropic $g$ factor with $g_c\,{>}\,g_{ab}$ can be described in terms of the trigonally elongated octahedron gives an insight into the highly anisotropic magnetic susceptibility $\chi$ and saturation magnetization $M_{\rm s}$ in the honeycomb-lattice quantum magnet $\alpha$-RuCl$_3$, where the RuCl$_6$ octahedron is trigonally compressed. We infer that the conditions ${\chi}^{ab}\,{\gg}\,{\chi}^{c}$ and $M_{\rm s}^{ab}\,{\gg}\,M_{\rm s}^{c}$ observed in $\alpha$-RuCl$_3$ are due to the condition of $g_{ab}\,{\gg}\,g_{c}$ owing to the trigonally compressed octahedron~\cite{Kubota}.

\begin{figure}[t]
\centering
\includegraphics[width=4.5cm,clip]{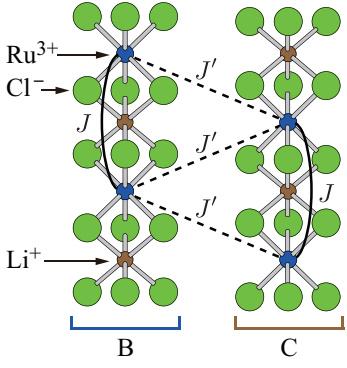}
\caption{(Color online) Exchange interaction $J$ in structural chains B and C, and interchain exchange interaction $J^{\prime}$ between these chains.}
\label{fig:exchange}
\end{figure}

\begin{figure}[t]
\centering
\includegraphics[width=8cm,clip]{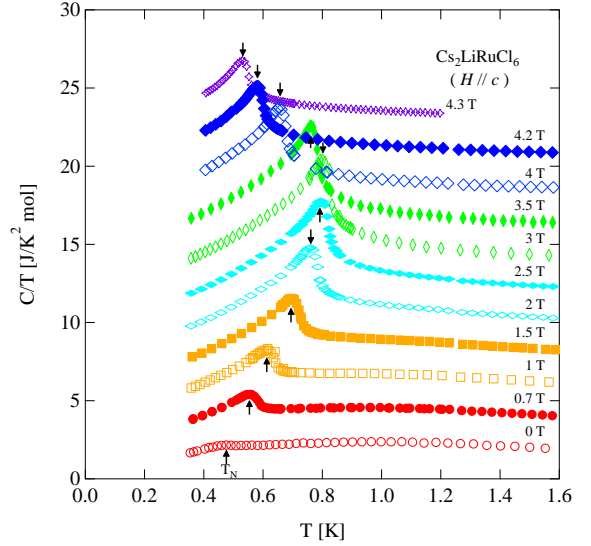}
\caption{(Color online) Total specific heat divided by temperature $C/T$ as a function of temperature measured at various magnetic fields for $H\,{\parallel}\,c$. Vertical arrows indicate magnetic phase transition temperatures $T_{\rm N}(H)$.}
\label{fig:heat_nonzero}
\end{figure}

The magnetization curves of Cs$_2$LiRuCl$_6$ shown in Fig.~\ref{fig:mag2}(b) are convex functions of the magnetic field up to the saturation and similar to the magnetization curve for the spin-1/2 Heisenberg antiferromagnetic chain calculated with $J/k_{\rm B}\,{=}\,3.7$ K~\cite{Griffiths}. 
The temperatures that give the maximum magnetic susceptibility $T_{\rm max}^{{\rm HAFC}}(\chi)$ and the maximum specific heat divided by temperature $T_{\rm max}^{{\rm HAFC}}(C/T)$ for the spin-1/2 Heisenberg antiferromagnetic chain are given by $T_{\rm max}^{{\rm HAFC}}(\chi)\,{=}\,0.64J/k_{\rm B}$ and $T_{\rm max}^{{\rm HAFC}}(C/T)\,{=}\,0.31J/k_{\rm B}$, respectively~\cite{Johnston}. In Cs$_2$LiRuCl$_6$, $T_{\rm max}^{{\rm exp}}(\chi)\,{\simeq}\,2.2$ K and $T_{\rm max}^{{\rm exp}}(C/T)\,{\simeq}\,1.0$ K. The value of $T_{\rm max}^{{\rm exp}}(\chi)/T_{\rm max}^{{\rm exp}}(C/T)$ is close to that of $T_{\rm max}^{{\rm HAFC}}(\chi)/T_{\rm max}^{{\rm HAFC}}(C/T)$.  From $T_{\rm max}^{{\rm exp}}(\chi)$ and $T_{\rm max}^{{\rm exp}}(C/T)$, the exchange constant is estimated to be $J/k_{\rm B}\,{\simeq}\,3.4$ K, which agrees approximately with the exchange constant of $J/k_{\rm B}\,{=}\,3.7$ K estimated from the average of the saturation fields. The ordering temperature $T_{\rm N}\,{=}\,0.48$ K is much lower than $J/k_{\rm B}\,{\simeq}\,3.4-3.7$ K.
From these results, we infer that Cs$_2$LiRuCl$_6$ can be approximately described as a coupled spin-1/2 Heisenberg-like antiferromagnetic chain. 

It is natural to assume that the structural chains ${-}$\,RuCl$_3$\,${-}$\,LiCl$_3$\,${-}$ in chains B and C are magnetic chains and that the path ${-}$\,Ru\,${-}$\,Cl\,${-}$\,Cl\,${-}$\,Ru\,${-}$ is the dominant path of the antiferromagnetic superexchange interaction $J$, as shown in Fig.~\ref{fig:exchange}. 
The interchain exchange interaction $J^{\prime}$ between structural chains B and C will be antiferromagnetic because the superexchange path is very similar to that in the hexagonal ABX$_3$ antiferromagnets~\cite{Collins}. As seen from Fig.~\ref{fig:exchange}, the intrachain and interchain exchange interactions cause spin frustration, which generally leads to an incommensurate spin structure in the ordered state. Because magnetic Ru$^{3+}$ ions are randomly distributed in structural chain A, it is difficult to construct its magnetic model.  


\begin{figure}[t]
\centering
\includegraphics[width=7.5cm,clip]{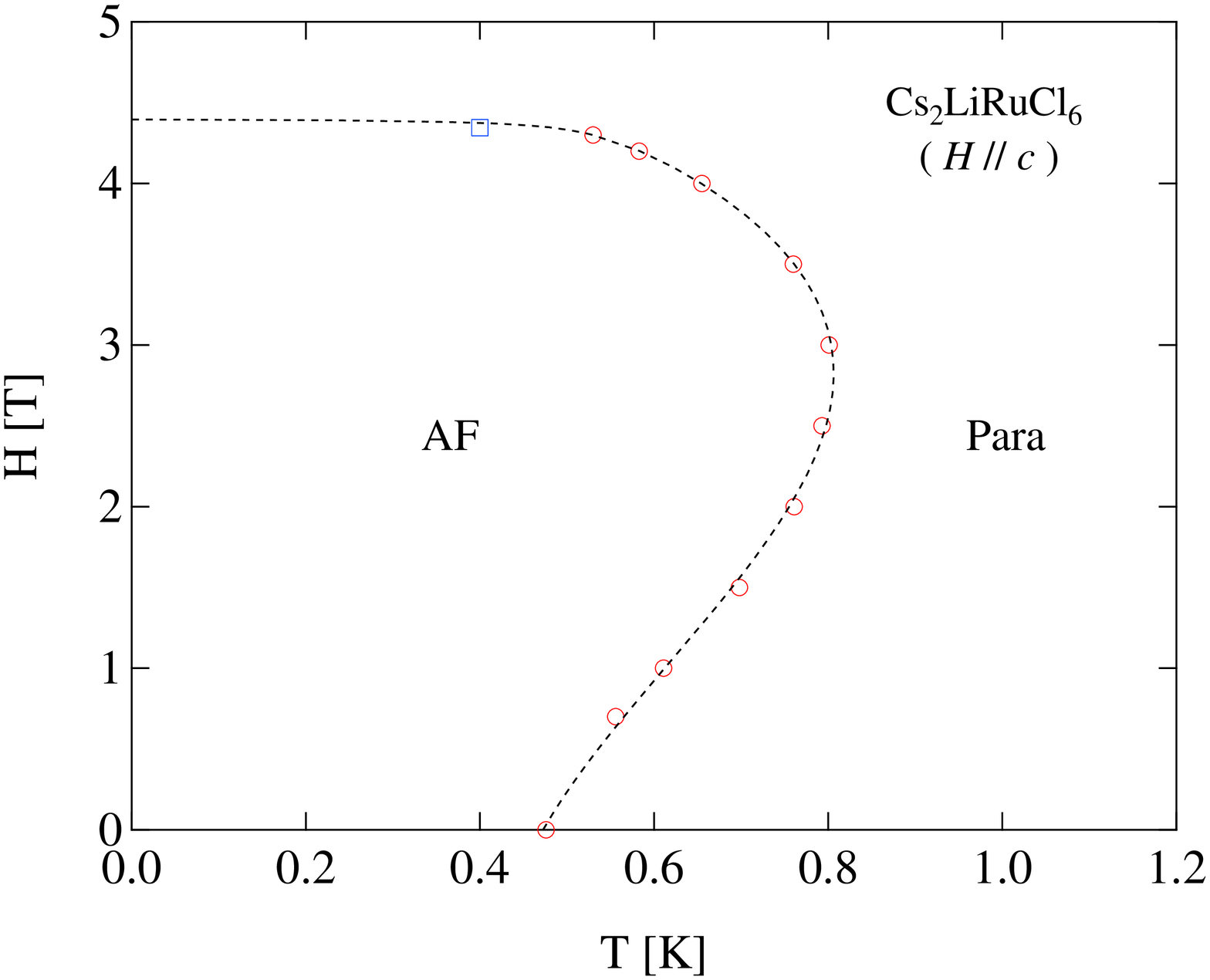}
\caption{(Color online) Magnetic field vs temperature phase diagram of Cs$_2$LiRuCl$_6$ for $H\,{\parallel}\,c$. Open circles and the square are transition points determined from the temperature dependence of $C/T$ and the saturation field at $T\,{=}\,0.4$ K, respectively. The dotted line is a guide to the eye.}
\label{fig:phase}
\end{figure}

To determine the magnetic phase diagram, we measured the specific heat in magnetic fields. Figure~\ref{fig:heat_nonzero} shows the temperature dependence of total specific heat divided by temperature $C/T$ measured at various magnetic fields for $H\,{\parallel}\,c$. With increasing magnetic field $H$, the transition temperature $T_{\rm N}(H)$ increases, and the $\lambda$-like anomaly of the specific heat is enhanced. $T_{\rm N}(H)$ reaches a maximum at $H\,{\simeq}\,3$ T then decreases. The transition points for $H\,{\parallel}\,c$ are summarized in Fig.~\ref{fig:phase}. The antiferromagnetic (AF) phase markedly protrudes into the paramagnetic (Para) phase. In the low-field region, $T_{\rm N}(H)$ increases rapidly with increasing magnetic field. We infer that this behavior arises from the suppression of the spin fluctuation by the magnetic field, which is characteristic of the quasi-one-dimensional Heisenberg-like antiferromagnets as observed in CuCl$_2\cdot$2NC$_5$H$_5$ with spin-1/2~\cite{Jonge,Tinus}, CsNiCl$_3$ with spin-1~\cite{Johnson3,Beckmann} and (CH$_3$)$_4$NMnCl$_3$ with spin-5/2~\cite{Jonge,Takeda}.

\section{Conclusion}
We have presented the results of structural analysis and magnetic measurements on Cs$_2$LiRuCl$_6$. The crystal structure is hexagonal $P6_322$. The structure consists of three kinds of chemical chain, A, B and C, composed of face-sharing RuCl$_6$ and LiCl$_6$ octahedra, as shown in Fig.~\ref{fig:Structure}. In chains B and C, RuCl$_6$ and LiCl$_6$ octahedra are arranged almost alternately, while in chain A, these octahedra are randomly arranged. The ordering of Ru$^{3+}$ and Li$^+$ in chains B and C, and their disordering in chain A can be mapped on the partially disordered ground state in the antiferromagnetic Ising model on the triangular lattice by making the lattice points occupied by Ru$^{3+}$ and Li$^+$ correspond to the Ising spins ${\sigma}^z\,{=}\,+1$ and $-1$, respectively.

The magnetic susceptibility and magnetization process in Cs$_2$LiRuCl$_6$ were found to be highly anisotropic, which is mainly owing to the anisotropy of the $g$ factor. The $g$ factors for magnetic fields parallel to the $c$ axis and $ab$ plane were determined by the electron paramagnetic resonance as $g_c\,{=}\,2.72$ and $g_{ab}\,{=}\,1.50$, respectively. The condition $g_c\,{>}\,g_{ab}$ can be attributed to the trigonally elongated RuCl$_6$ octahedron in Cs$_2$LiRuCl$_6$. This gives an insight into the anisotropic magnetic properties observed in $\alpha$-RlCl$_3$, which is a candidate material following the Kitaev model. The magnetization curves in Cs$_2$LiRuCl$_6$ for $H\,{\parallel}\,c$ and $H\,{\parallel}\,ab$ exhibit a reasonably sharp saturation anomaly and roughly coincide when normalized by the $g$ factor. This indicates that the magnetic anisotropy and the Kitaev term are relatively small as compared with the dominant Heisenberg term $J$. From the magnetic and thermodynamic properties, it was found that Cs$_2$LiRuCl$_6$ can be described as a coupled spin-1/2 Heisenberg-like antiferromagnetic chain.  We determined the magnetic phase diagram for $H\,{\parallel}\,c$ as shown in Fig.~\ref{fig:phase}. In the low-field region, the ordering temperature $T_{\rm N}(H)$ increases rapidly with increasing magnetic field. The suppression of the spin fluctuation by the magnetic field is considered to be the origin of this behavior.

\section*{Acknowledgment}
This work was supported by Grants-in-Aid for Scientific Research (A) (No.~17H01142) and (C) (No.~19K03711) from Japan Society for the Promotion of Science. This work was performed under the Inter-University Cooperative Research Program of the Institute for Materials Research, Tohoku University (Proposal No. 19K0068).

\end{document}